\documentstyle[12pt]{article}
\begin{document}
\renewcommand{\thepage}{ }
\begin{titlepage}
\title{
\hfill
\parbox{4cm}{\normalsize KUNS 1220\\HE(TH)93/10\\cond-mat/9309018\\
September 1993}\\
\vspace{1.5cm}
Topological Landau-Ginzburg Theory \\
for Vortices in Superfluid $^4$He}
\author{
M.\ Hatsuda$^{1,2}$,  S.\ Yahikozawa$^1$, P.\ Ao$^2$
and D.\ J.\ Thouless$^2$  \\
{}\\
$^1${\normalsize\em Department of Physics, Kyoto University,
Kyoto 606-01, Japan} \\
$^2${\normalsize\em Physics Department, FM-15,
University of Washington, Seattle, WA 98195, USA}}
\date{}
\maketitle
\begin{abstract}
\normalsize
We propose a new Landau-Ginzburg theory
for arbitrarily shaped vortex strings in superfluid $^4$He.
The theory contains a topological term and
directly describes vortex dynamics.
We introduce gauge fields in order to remove singularities
from the Landau-Ginzburg order parameter of the superfluid,
so that two kinds of gauge symmetries appear,
making the continuity equation and
conservation of the total vorticity manifest.
The topological term gives rise to
the Berry phase term in  the vortex mechanical actions.
\end{abstract}
\end{titlepage}

\newpage
\renewcommand{\thepage}{\arabic{page}}
\setcounter{page}{1}
\baselineskip=17pt plus 0.2pt minus 0.1pt

Since the existence of quantized vortices was predicted by
Onsager and Feynman, vortices have been observed
in superfluid $^4$He and $^3$He, and in superconductor systems.
At low temperature in superfluid helium the quantized vortex obeys
the classical hydrodynamical law
that the vortex moves with the local velocity of the fluid,
while the vorticity quantization
comes from the fact that the superfluid is a quantum state
described by a wavefunction. The vortex
dynamics is governed by classical hydrodynamics and the quantum
aspects of the system
is governed by a non-linear Schr$\ddot{\rm o}$dinger
equation which is  equivalent to the Landau-Ginzburg theory for
superfluid developed by Ginzburg, Pitaevskii
and Gross (GPG) \cite{GPita}.
By inserting a suitable form for the phase of the field
by hand on a case-by-case basis it has been shown
that a Landau-Ginzburg theory produces vortex dynamics \cite{Ld,CMK}.
However there is no satisfactory theory which describes
both vortex dynamics and quantum properties of the vortex.

In this paper we propose a topological Landau-Ginzburg theory
for vortices in superfluid $^4$He. The characteristic features
of our formalism are as follows.
(i) We introduce  a gauge field, $A_{\mu}$, in the GPG theory
in such a way that
$A_{\mu}$ carries  the singularities in the phase of
the Landau-Ginzburg order parameter: the phase therefore
becomes single valued.
In order not to change physical observables we introduce
$A_{\mu}$ gauge covariantly, and we choose the
condition that the dual field strength of
$A_{\mu}$  coincides with the vorticity tensor.
This condition is imposed by
using a rank two antisymmetric tensor Lagrange multiplier,
$B_{\mu\nu}$. There are
two kinds of gauge symmetries, which lead to
the continuity equation
and to the conservation of the total vorticity.
(ii) A topological term, ``BF term''
($\varepsilon^{\mu\nu\rho\lambda}
B_{\mu\nu}F_{\rho\lambda}$), where $F_{\mu\nu}$ is the field strength,
$F_{\mu\nu} =\partial_{\mu}A_{\nu}-\partial_{\nu}A_{\mu}$,
and a coupling term,  $B_{\mu\nu}J^{\mu\nu}$,
to the vorticity tensor
$J^{\mu\nu}$ are required to reproduce the Berry phase term.
Because the BF term does not couple
to the 3+1 dimensional metric,
it is a so-called topological term.
In general the BF term is used in evaluating linking numbers
which are topological numbers counting
how many times a string and a membrane
are entangled in 3+1 dimensions \cite{BBRT},
while the Berry phase term
in the vortex mechanical action is similar to the Hopf term
which counts the instanton number in the $O(3)$
nonlinear sigma model \cite{Wz83}.
The topological BF term is a generalization
of the Chern-Simons term which plays an important role
in the study of the fractional quantized Hall effect
and anyon systems in 2+1 dimensions \cite{Wz83}.
It is desirable to include such a topological term since
a vortex is a topological excitation in the sense that
it is not obtained by a continuous
deformation from the ground state.
(iii) The vorticity tensor,
whose time components, $J^{0i}$,  correspond to a vorticity vector,
has general form so that it can describe arbitrarily shaped
vortex strings or rings.  Regularization, if needed, involves
regularizing only this tensor, so that the density $\rho$ and
the velocity, {\boldmath $v$}, never become singular.
(iv) Since the vorticity tensor contains vortex coordinates
explicitly, this action directly leads to the equation of motion of
vortices as well as the field equations
for the order parameter. This action also reproduces
the correct vortex mechanical action which contains the Berry
phase term.

In connection with point (ii),
the rank two anti-symmetric tensor field $B_{\mu\nu}$
(Kalb-Ramond field) has been used to describe the vortex dynamics
in superfluid \cite{LR}. However the theory used was
Kalb-Ramond theory  which completely
differs from our approach.
In Kalb-Ramond theory \cite{KbR}, the action
contains the square of the field strength
for $B_{\mu\nu}$, $H_{\mu\nu\rho}H^{\mu\nu\rho}$,
so that $B_{\mu\nu}$
becomes dynamical and propagates in space-time. The essential
difference is that the form of $B_{ij}$ is set by hand
to be $B_{ij}=\epsilon_{ijk}x^k$ in the previous approach, while
the condition $B_{ij}=\epsilon_{ijk}x^k$
appears naturally in our approach as we show later.
Therefore our formalism is
perhaps more natural when describing the vortex
dynamics in a superfluid.

One of the motivations of this work is how to treat the nucleation
of quantized vortices. Among the most interesting problems
in vortex physics are the nucleation and annihilation
of quantized vortices
and the mechanism of breakdown of the superfluidity
and the superconductivity through the vortex nucleation\cite{MVD}.
Nucleation involves the creation of a vortex from nothing, and
annihilation destroys an existing vortex ring.
Although recently interesting classical numerical simulations of
vortex nucleation have been reported \cite{FPR},
quantum field theory is required for a complete treatment
of such processes.
However, for a theory of these processes
and for other problems involving vortices, we need to
develop a suitable parametrization in which
the slowly changing coordinates, such as the path of a vortex ring,
separate easily from the high frequency modes.
Then by taking into account suitable instanton-like solutions of the
low frequency sector we can calculate the nucleation rates or other
quantities of physical interest.
We hope that our new theory will be useful for this purpose.

We begin by proposing a gauged Landau-Ginzburg theory with a
topological term
\begin{eqnarray}
&&S=\int\! d^4x \biggl[\hbar\psi^*\left(i\partial_0 + A_0\right)\psi
- \frac{\hbar^2}{2m}\left\vert\left(i\partial_i+ A_i
\right)
\psi\right\vert^2
- V(\psi)  \nonumber  \\
&&\hspace*{30mm}
+\frac{\hbar}{2m}\varepsilon^{\mu\nu\rho\sigma}B_{\mu\nu}
F_{\rho\sigma} + B_{\mu\nu}J^{\mu\nu}\biggr],
\label{eq.1} \\
&&J^{\mu\nu}(x)=\sum_{a=1}^{N}\gamma_a \int\!d\tau  d\sigma
\frac{\partial X_a^{[\mu}}
{\partial \tau} \frac{\partial X_a^{\nu ]}}{\partial \sigma}
\delta^{(4)}\left(x-X_a(\tau, \sigma)\right),
\label{eq.2}
\end{eqnarray}
where $\varepsilon^{\mu\nu\rho\sigma}$
is the complete antisymmetric tensor and $V(\psi)$ is a
potential which is a gauge-invariant function of $\psi$.
In general, one assumes that
$V(\psi)=\lambda(\vert\psi\vert^2-\rho_0)^2$
so that the number density $\vert\psi\vert^2$ does not deviate
too much from $\rho_0$ which is a constant.
$J^{\mu\nu}$ is the vorticity tensor
representing vortex strings or rings
on which singularities appear while
$X_a^{\mu}$ and $\gamma_a$ are the vortex position and the
vorticity of the $a$-th vortex respectively, and
$A^{[\mu}B^{\nu]}\equiv A^{\mu}B^{\nu}-A^{\nu}B^{\mu}$.
The condition that the wavefunction  be single valued results
in quantization of $\gamma_a$;
$\gamma_a=(h/m) n_a$ where $n_a$ is some integer.
Because of the form of the vorticity tensor, the variation of
the action with respect to the vortex coordinates $X_a^{\mu}$
gives the equation of motion of the $a$-th vortex directly.
Since singularities appear only through the vorticity tensor,
it is possible to regularize, if necessary, by modifying the form
of the vorticity tensor, for example the $\delta$-function can
be replaced by some other distribution.

An important property of our action is that
there are two gauge symmetries. The first one
is a usual gauge symmetry;
$A_{\mu}\rightarrow A_{\mu}+\partial_{\mu}\Lambda$
and $\psi \rightarrow e^{i\Lambda}\psi$ with a regular function
$\Lambda$, which leads to the continuity equation
${\partial}{\rho}/{\partial} t +
\nabla\!\cdot\!(\rho \mbox{\boldmath $v$})=0$.
The second is given by
$B_{\mu\nu}\rightarrow B_{\mu\nu}+\partial_{\mu}\Lambda_{\nu}
-\partial_{\nu}\Lambda_{\mu}$, where
$\Lambda_{\mu}$ is also a regular function.
The corresponding conserved current is $J^{\mu\nu}$,
so that the total vorticity $\int\! d^3x {J^{0i}}$
is conserved.

If we consider three dimensional vortex rings, the trajectory of
a vortex ring, which is a sheet, may be parametrized
by $\tau$ and $\sigma$.
Since the trajectory of a vortex ring is a sheet,
the vorticity tensor which describes
vortex trajectories becomes a rank two tensor.
On the other hand, for the trajectory of a point
vortex it becomes a vector.
In order to describe non-relativistic situation
we can identify $\tau$ and $X_a^0$ with
the time axis, that is, $\tau=X_a^0=t$.
We may rewrite $J^{\mu\nu}$ in terms of two vectors
${\mbox{\boldmath $J$}}=(J^{0i})$ and
$\mbox{\boldmath $j$}=(\frac{1}{2}\epsilon^{ijk}J^{jk})$,
and {\boldmath $J$} just coincides with an usual vorticity vector
\begin{eqnarray}
\mbox{\boldmath $J$}(\mbox{\boldmath $x$})&=&
\sum_{a=1}^{N}\gamma_a \oint_{\Gamma_a}\! d\mbox{\boldmath $X$}_a
\delta^{(3)}
\left(\mbox{\boldmath $x$}-\mbox{\boldmath $X$}_a(t,\sigma)\right)
\label{eq.3},  \\
\mbox{\boldmath $j$}(\mbox{\boldmath $x$})&=&
\sum_{a=1}^{N}\gamma_a \oint_{\Gamma_a}\!
\dot{\mbox{\boldmath $X$}}_a \times d\mbox{\boldmath $X$}_a
\delta^{(3)}
\left(\mbox{\boldmath $x$}-\mbox{\boldmath $X$}_a(t,\sigma)\right),
\label{eq.4}
\end{eqnarray}
where $\Gamma_a$ is a ring configuration of the $a$-th vortex ring.
The conservation law of $J^{\mu\nu}$ written in terms of
these vectors becomes $\nabla \cdot \mbox{\boldmath $J$}=0$ and
$\partial \mbox{\boldmath $J$}/\partial t=
\nabla \times \mbox{\boldmath $j$}$.

An important feature of our theory is that the equation
of motion of vortices can be obtained from our action directly.
The variation of the action with respect to the vortex coordinate
$X_a^i$ gives
\begin{eqnarray}
H_{0ij}\frac{\partial X_a^{j}}{\partial \sigma}-
H_{ijk}\frac{\partial X_a^{j}}
{\partial t} \frac{\partial X_a^{k}}{\partial  \sigma}=0 ,
\label{eq.5}
\end{eqnarray}
where $H_{\mu\nu\rho}=\partial_{\mu}B_{\nu\rho}+
\partial_{\nu}B_{\rho\mu}
+\partial_{\rho}B_{\mu\nu}$, and $H_{0ij}$ and $H_{ijk}$ are
determined by the equations of motion of $A_0$ and $A_i$ to be
\begin{eqnarray}
&\displaystyle H_{ijk}=-\frac{m\rho}{2} \epsilon_{ijk} ,&
\label{eq.6}  \\
&\displaystyle H_{0ij}=\frac{m\rho}{2} \epsilon_{ijk}v^k ,&
\label{eq.7}
\end{eqnarray}
where   $\mbox{\boldmath $v$}=(\hbar /m)
(\nabla\theta+\mbox{\boldmath $A$})$.
Substituting equations (\ref{eq.6}) and (\ref{eq.7})
into (\ref{eq.5}), we obtain the equation of motion of the vortex as
\begin{eqnarray}
\frac{\partial \mbox{\boldmath $X$}_a}{\partial t}=
\mbox{\boldmath $v$}(\mbox{\boldmath $X$}_a)+
\alpha_a\frac{\partial \mbox{\boldmath $X$}_a}
{\partial \sigma},
\label{eq.8}
\end{eqnarray}
where $\alpha_a$ are arbitrary coefficients reflecting the
reparametrization freedom of $\sigma$: there is
a parametrization of $\sigma$ such that the last term
in (\ref{eq.8}) vanishes. The vortex equation of motion
is the same as the equation obtained in classical
hydrodynamics, and specifies that vortices move with the local fluid
velocity. Note that we have also shown that the equation
$\partial\mbox{\boldmath $X$}_a/\partial t=
\mbox{\boldmath $v$}(\mbox{\boldmath $X$}_a)$
holds not only in an incompressible superfluid but also in
a compressible superfluid, since we did not assume
any condition such as $\rho=\rho_0$ in the above derivation.

As well as the equation of motion for vortices, we have
following field equations,
obtained by varying the density $\rho (x)$
which is $\vert \psi \vert^2$ and the phase of $\psi$, $\theta (x)$
\begin{eqnarray}
&\displaystyle \hbar\dot\theta -
\hbar A_0 + \frac{m}{2}\mbox{\boldmath $v$}^2
-\frac{\hbar^2}{2m}
\frac{1}{\sqrt{\rho}}\Delta\sqrt{\rho}
+\frac{\partial V}{\partial {\rho}} =0 ,&
\label{eq.9}  \\
&\displaystyle \frac{\partial\rho}{\partial t}+
\nabla\!\cdot\!(\rho\mbox{\boldmath $v$})=0 .&
\label{eq.10}
\end{eqnarray}
The first equation is similar to the Bernoulli theorem
and second is the  continuity equation.
The field equation for $\rho$ gives the dependence
of $\rho$ on the position of the vortex ring $\mbox{\boldmath $X$}_a$.
In general, $\rho(\mbox{\boldmath $x$};
\{\mbox{\boldmath $X$}_{a'}\})=\rho_s(\mbox{\boldmath $x$})+
\delta\rho(\mbox{\boldmath $x$},
\mbox{\boldmath $X$}_a)$ where $\rho_s(\mbox{\boldmath $x$})$
is the density in the absence
of the $a$-th vortex ring and
$\delta\rho(\mbox{\boldmath $x$},\mbox{\boldmath $X$}_a)$ is
the modification due to the presence of the $a$-th vortex ring.
Indeed $\delta\rho(\mbox{\boldmath $x$},\mbox{\boldmath $X$}_a)$
contains the contribution from the core of the $a$-th vortex ring.
Variations with respect to $B_{i0}$ and $B_{ij}$ lead to
\begin{eqnarray}
&\displaystyle \nabla \times \mbox{\boldmath $A$} =
\frac{m}{\hbar}\mbox{\boldmath $J$}  ,&
\label{eq.11}  \\
&\displaystyle \frac{\partial\mbox{\boldmath $A$}}{\partial t} +
\nabla A^0 =\frac{m}{\hbar} \mbox{\boldmath $j$}  .&
\label{eq.12}
\end{eqnarray}
The first equation is a constraint which restricts
the space component of the gauge field
and the second equation determines
the time component of the gauge field.

Now we will show that our new action reproduces
the vortex mechanical action.
Since this theory is a macroscopic theory,
there is a cutoff of the theory which must be bigger than
the atomic scale $\sim\!\mbox{\AA}$ of the underlying microscopic
dynamics. Since the coherence length
of the vortex core in the superfluid, $\xi$, is of the order of one
angstrom which is comparable with the cutoff of this theory,
we neglect contributions from vortex cores which are
$O(\xi^3 /L^3)$ with a container size $L$.
We assume that time and spatial derivatives of
{\boldmath $X$} are small.
We also assume incompressibility, $\rho(x)=\rho_0=\mbox{constant}$,
and that the phase of $\psi$, $\theta$, be zero for simplicity,
so that the velocity {\boldmath $v$}
almost coincides with $(\hbar /m)\mbox{\boldmath $A$}$
except at the vortex cores.
According to the constraint (\ref{eq.6})
which is $\nabla\cdot\mbox{\boldmath $b$}(\mbox{\boldmath $x$}) =
-(m/2)\rho (\mbox{\boldmath $x$})$,
the integrand of
$\int\!d^4x \mbox{\boldmath $b$} \cdot
\partial\mbox{\boldmath $A$}/\partial t$,
which is a part of the BF term, is a total time derivative
except near the vortex cores, so that this contribution
is also neglected.
Using the conditions (\ref{eq.6}) and (\ref{eq.11})
with the above assumptions, then the original action (\ref{eq.1})
reduces to
\begin{eqnarray}
S=\int\! d^4x\biggl[ 2\mbox{\boldmath $b$}\cdot\mbox{\boldmath $j$}-
\frac{\hbar^2 \rho_0}{2m} \mbox{\boldmath $A$}^2
\biggr] ,
\label{eq.13}
\end{eqnarray}
where $\mbox{\boldmath $b$} =(\frac{1}{2} \epsilon^{ijk} B^{jk})$ and
$ 2\mbox{\boldmath $b$}\cdot\mbox{\boldmath $j$}=B_{ij}J^{ij}$.
Imposing gauge fixing conditions
$\nabla \cdot \mbox{\boldmath $A$}=0$ and
$\nabla \times \mbox{\boldmath $b$}=0$ and solving
(\ref{eq.11}) and (\ref{eq.6}),
we obtain
\begin{eqnarray}
\mbox{\boldmath $A$}(\mbox{\boldmath $x$})&=&
\frac{m}{2h}\sum_{a=1}^{N}\gamma_a\oint\! d\mbox{\boldmath $X$}_a\times
\frac{\mbox{\boldmath $x$}-\mbox{\boldmath $X$}_a}
{\left\vert \mbox{\boldmath $x$}-\mbox{\boldmath $X$}_a \right\vert^3} ,
\label{eq.14}  \\
\mbox{\boldmath $b$}(\mbox{\boldmath $x$})&=&
-\frac{\rho_0 m}{6} \mbox{\boldmath $x$}.
\label{eq.15}
\end{eqnarray}
Here we stress that these solutions are fully determined by the
theory.

We substitute {\boldmath $A$} and {\boldmath $b$}
into the action (\ref{eq.13}),
which becomes
\begin{eqnarray}
S=\rho_0 m \int\! dt\biggl[ \frac{1}{3}\sum_{a=1}^{N}
\gamma_a\oint\!
d\mbox{\boldmath $X$}_a\cdot(\dot{\mbox{\boldmath $X$}}_a
\times\mbox{\boldmath $X$}_a)
-\frac{1}{8\pi}\sum_{a,b} \gamma_a\gamma_b
\oint\!\oint\! \frac{d\mbox{\boldmath $X$}_a
\cdot d\mbox{\boldmath $X$}_b}{\left\vert\mbox{\boldmath $X$}_a
-\mbox{\boldmath $X$}_b\right\vert} \biggr].
\label{eq.16}
\end{eqnarray}
This is of the same form as the action of a vortex ring
in an incompressible perfect fluid \cite{RR}.
If we consider an adiabatic process in which the final configuration
coincides with the initial configuration,
the first term in (\ref{eq.16}) can be interpreted as the Berry phase
\cite{Ber}, which is $-i\hbar\oint\! dt\left\langle\Psi(X_a(t))\vert
d\Psi(X_a(t))/dt\right\rangle$
where $\Psi$ is a microscopic Feynman type wavefunction.
The second term in (\ref{eq.16}), which was a kinetic term
of the local fluid corresponding to
$-(\hbar^2\rho_0 /2m)\int\! d^4x \mbox{\boldmath $A$}^2$
in (\ref{eq.13}), represents the interaction
between vortex rings. Thus our new
theory naturally includes the vortex dynamics. Note that
by using field equations there is an interesting relation
$  \int\! d^4 x B_{ij}J^{ij}
=\hbar\int\! d^4x \rho A_0$: it is clear that the left
hand side gives the phase change when the vortex ring
is moved around a closed trajectory, that is the Berry phase,
but it is not obvious that the right hand side also does.

It is easy to get the vortex mechanical action in the
compressible case.
For example, the term corresponding to the Berry phase term
$ \int\! d^4 x 2\mbox{\boldmath $b$}\cdot\mbox{\boldmath $j$}=
\int\! d^4 x B_{ij}J^{ij}$
is given by
\begin{eqnarray}
\frac{m}{4\pi}\sum_{a=1}^{N}\gamma_a\int\! d^4x
\oint\! d\mbox{\boldmath $X$}_a\cdot
\Bigl(
\dot{\mbox{\boldmath $X$}}_a\times
\frac{\mbox{\boldmath $x$}-\mbox{\boldmath $X$}_a}
{\left\vert \mbox{\boldmath $x$}-\mbox{\boldmath $X$}_a \right\vert^3}
\Bigr) \rho(x).
\label{eq.17}
\end{eqnarray}
The Berry phase for a compressible case
can be calculated by using this formula.

We now proceed to discuss more familiar
two dimensional case in which
point vortices move in a thin superfluid film
which lying vertical to the $z$ axis.
The vortex string is directed to $z$ axis
and has no $z$-coordinate dependence, so that $\sigma$ is
identified as $\sigma =X_a^3$.
The vorticity tensor in two dimensions is then given
\begin{eqnarray}
\mbox{\boldmath $J$}(\mbox{\boldmath $r$}) &=&
\sum_{a=1}^{N}\gamma_a \mbox{\boldmath $e$}_z \delta^{(2)}
\left(\mbox{\boldmath $r$}-\mbox{\boldmath $R$}_a(t)\right),
\label{eq.18}  \\
\mbox{\boldmath $j$}(\mbox{\boldmath $r$})&=&
\sum_{a=1}^{N}\gamma_a\dot{\mbox{\boldmath $R$}}_a
\times \mbox{\boldmath $e$}_z
\delta^{(2)}
(\mbox{\boldmath $r$}-\mbox{\boldmath $R$}_a(t)),
\label{eq.19}
\end{eqnarray}
where $\mbox{\boldmath $e$}_z$ is a unit vector along the $z$ axis,
{\boldmath $r$} is the two dimensional coordinate and
$\mbox{\boldmath $R$}_a$ is
the two dimensional position of the $a$-th point vortex.
Solving the constraint conditions (\ref{eq.6}) and (\ref{eq.11})
analogous to three dimensional case with same conditions in the three
dimensional case, we get
\begin{eqnarray}
\mbox{\boldmath $A$}(\mbox{\boldmath $r$})&=&
\frac{m}{h}\sum_{a=1}^{N}\gamma_a \mbox{\boldmath $e$}_z \times
\frac{\mbox{\boldmath $r$}-\mbox{\boldmath $R$}_a(t)}
{\left\vert\mbox{\boldmath $r$}-\mbox{\boldmath $R$}_a(t)\right\vert} ,
\label{eq.20}  \\
\mbox{\boldmath $b$}(\mbox{\boldmath $r$})&=&
-\frac{m\rho_0}{4} \mbox{\boldmath $r$}.
\label{eq.21}
\end{eqnarray}
As a result the action for point vortices in incompressible
two dimensional superfluid becomes
\begin{eqnarray}
S_2=\rho_0 m\int\! dt \biggl[\frac{1}{2}\sum_{a=1}^{N}\gamma_a
{\mbox{\boldmath $e$}}_z\cdot (\dot{\mbox{\boldmath $R$}}_a
\times\mbox{\boldmath $R$}_a)
+\frac{1}{4\pi}
\sum_{a\neq b}\gamma_a\gamma_b\ln\vert\mbox{\boldmath $R$}_a
-\mbox{\boldmath $R$}_b\vert\biggr].
\label{eq.22}
\end{eqnarray}
Taking into account the equation of motion, we can check
that the velocity of the vortex $\mbox{\boldmath $R$}_a$ coincides
with the local velocity field at $\mbox{\boldmath $R$}_a$ \cite{HalW}.

It is interesting to formulate 2+1 dimensional theory
by dimensional reduction of the original 3+1 dimensional theory:
assuming $z$-independence of the fields $\psi$,
$A_{\mu}$ and $B_{\mu\nu}$ and $A_3 =0$, surviving terms
$\varepsilon^{\mu\nu\rho\sigma}
B_{\mu\nu}F_{\rho\sigma}$ and $B_{\mu\nu}J^{\mu\nu}$ in
$3+1$ dimensional action (\ref{eq.1}) become
$-2 \epsilon^{\mu\nu\rho}B_{\mu}F_{\nu\rho}$
and $2 B_{\mu}J^{\mu}$ in $2+1$ dimensional action respectively
where $B_{\mu}\equiv B_{3\mu}$
and $J^{\mu}\equiv J^{3\mu}$, $\mu=0,1,2$.
Using this 2+1 dimensional BF term,
we can also obtain the vortex dynamics.

Now let us compare our formalism with that of the GPG theory.
The GPG action is given by
\begin{eqnarray}
S_{{\rm GPG}}=\int\! d^4x\biggl[i\hbar\phi^*\partial_0\phi
-\frac{\hbar^2}{2m}
\left\vert\nabla\phi\right\vert^2-\lambda(\left\vert\phi\right\vert^2
-\rho_0)^2\biggr],
\label{eq.23}
\end{eqnarray}
where $\phi$ is a complex Bose field which is allowed to have
a singular phase. The phase $\theta_{{\rm GPG}}$ of $\phi$
includes both
a regular part and a singular part which is multi-valued:
\begin{eqnarray}
\theta_{{\rm GPG}}=\theta_{{\rm reg}}+\theta_{{\rm sing}}.
\label{eq.24}
\end{eqnarray}
The singular phase $\theta_{{\rm sing}}$
is the sum of the solid angles subtended by the vortex rings
\begin{eqnarray}
\theta_{{\rm sing}}=\frac{1}{2}\sum_{a=1}^{N}n_a
\int_{S_a}d\mbox{\boldmath $S$}'_a
\cdot\nabla\frac{1}{\left\vert\mbox{\boldmath $x$}
-\mbox{\boldmath $x$}'\right\vert},
\label{eq.25}
\end{eqnarray}
where $S_a$ is any surface bounded by the $a$-th vortex ring,
$\partial S_a =\Gamma_a$ \cite{Ld}. An important property of
$\theta_{{\rm sing}}$
is the noncommutativity of the differential on it
\begin{eqnarray}
(\partial_{\mu}\partial_{\nu}-\partial_{\nu}
\partial_{\mu})\theta_{{\rm sing}}
=-\frac{m}{2\hbar}\varepsilon_{\mu\nu\rho\sigma}J^{\rho\sigma}.
\label{eq.26}
\end{eqnarray}
If we take $-\partial_{\mu}\theta_{{\rm sing}}$
as the gauge field $A_{\mu}$,
the equation (\ref{eq.26}) becomes the same as the constraint
condition which is given by varying our action
(\ref{eq.1}) with respect to $B_{\mu\nu}$.
Furthermore, substituting $e^{i\theta_{{\rm sing}}}
\phi_{{\rm reg}}$ into $\phi$ we obtain
$\phi^*\partial_0\phi=\phi_{{\rm reg}}^*(\partial_0-iA_0)
\phi_{{\rm reg}}$ and
$\left\vert\nabla\phi\right\vert^2=
\left\vert(\nabla-i\mbox{\boldmath $A$})
\phi_{{\rm reg}}\right\vert^2$.
Therefore, it turns out that if one solves the field equations of
$A_{\mu}$ and $B_{\mu\nu}$ in our formalism and our field
$\psi$ is taken as $\phi_{{\rm reg}}$, then our action (\ref{eq.1})
becomes equivalent to the GPG action (\ref{eq.23})
whose phase is
$\theta_{{\rm GPG}}=\theta_{{\rm reg}}+\theta_{{\rm sing}}$
in the case where
$V(\psi)=\lambda(\left\vert\psi\right\vert^2-\rho_0)^2$.
Our formalism, however, has some advantages as previously mentioned.
Since the dependence of the vortex coordinates appears only in
$J^{\mu\nu}$, ours is useful to investigate the vortex dynamics.
The topological BF term and the gauge invariance play important roles.

Towards the nucleation of the vortex, next step is to find
an instanton-like solution which makes action finite.
As for future problems, the statistics of three dimensional
vortex rings is interesting because
it is directly related to linking number. Our theory
can be applied to the superconductor in which the contribution
from vortex cores may be important since the coherent length
is relatively large.
Applications to the quantum Hall system are also interesting.

The topological BF term appears in various areas of theoretical
physics, for example, it is induced by one loop effects
in a model with an anomalous $U(1)$
charge in superstring theory \cite{LNS} and
it is also used in a gravity theory
which is called 2-form gravity \cite{CDJM}. Therefore it is interesting
to consider some connections between such theories
and the present theory on the vortex in the superfluid.

\vspace{1cm}
\centerline{\bf Acknowledgements}
M. H. would like to thank the condensed matter group
for their hospitality during her stay at University of Washington.
She also expresses her thanks  to T. Hatsuda for  helpful discussions
and encouragements.  S. Y. would like to acknowledge R. Ikeda, T. Muto,
and M. Sato for useful discussions.
We are grateful to M. G. Mitchard for reading the manuscript carefully.
S. Y. was supported in part by Grant-in-Aid for Scientific Research
from Ministry of Education, Science and Culture
(\# 05230037 and 05740175).  P. A. and D. T. were
supported  by NSF grant \# DMR-9220733.

\end{document}